\documentclass[fleqn,usenatbib]{mnras}

\usepackage{newtxtext,newtxmath}

\usepackage[T1]{fontenc}

\usepackage{graphicx}	
\usepackage{amsmath}	
\usepackage{enumitem}
\usepackage{subfigure}
\usepackage{booktabs}
\usepackage{threeparttable}
\usepackage{xcolor}
\usepackage{ulem}
\usepackage{calc}
\usepackage{soul}
\usepackage{ftnxtra}
\usepackage{fnpos}

\newcommand{\phantomcode}{{\sc Phantom}}
\newcommand{\hd}{HD~143006}

\title{\hd{}: circumbinary planet or misaligned disc?}

\author[G. Ballabio et al.]{\parbox{\textwidth}{
G.\,Ballabio$^{1,2}$\thanks{E-mail: g.ballabio@qmul.ac.uk}, R.\,Nealon$^{2,3,4}$, R.\,D.\,Alexander$^2$, N.\,Cuello$^5$, C.\,Pinte$^{5,6}$ and D.\,J.\,Price$^6$ 
} \vspace{0.2cm}\\
$^{1}$ Astronomy Unit, School of Physics and Astronomy, Queen Mary University of London, Mile End Road, London, UK \\
$^{2}$ School of Physics and Astronomy, University of Leicester, Leicester, LE1 7RH, UK \\
$^{3}$ Centre for Exoplanets and Habitability, University of Warwick, Coventry CV4 7AL, UK \\
$^{4}$ Department of Physics, University of Warwick, Coventry CV4 7AL, UK \\
$^{5}$ Univ. Grenoble Alpes, CNRS, IPAG, F-38000 Grenoble, France \\
$^{6}$ School of Physics and Astronomy, Monash University, Clayton Vic 3800, Australia \\
}

\date{Accepted 2021 March 29. Received 2021 March 3; in original form 2021 January 25}

\pubyear{2021}

\begin{document}
\label{firstpage}
\pagerange{\pageref{firstpage}--\pageref{lastpage}}
\maketitle

\begin{abstract}
Misalignments within protoplanetary discs are now commonly observed, and features such as shadows in scattered light images indicate departure from a co-planar geometry. VLT/SPHERE observations of the disc around \hd{} show a large-scale asymmetry, and two narrow dark lanes which are indicative of shadowing. ALMA observations also reveal the presence of rings and gaps in the disc, along with a bright arc at large radii. We present new hydrodynamic simulations of \hd{}, and show that a configuration with both a strongly inclined binary and an outer planetary companion is the most plausible to explain the observed morphological features. We compute synthetic observations from our simulations, and successfully reproduce both the narrow shadows and the brightness asymmetry seen in IR scattered light. Additionally, we reproduce the large dust  observed in the mm continuum, due to a 10 Jupiter mass planet detected in the CO kinematics. 
Our simulations also show the formation of a circumplanetary disc, which is misaligned with respect to the outer disc. The narrow shadows cast by the inner disc and the planet-induced ``kink'' in the disc kinematics are both expected to move on a time-scale of $\sim$5--10~years, presenting a potentially observable test of our model. If confirmed, \hd{} would be the first known example of a circumbinary planet on a strongly misaligned orbit. 
\end{abstract}




\section{Introduction}\label{sec:intro}
High angular resolution observations at multiple wavelengths have shown a wealth of sub-structures within protoplanetary discs, such as ring-like features, horseshoes, spirals and other azimuthal asymmetries \citep[e.g.][]{2015ApJ...808L...3A, 2020ARA&A..58..483A, 2020ApJ...892..111F}, which are detected in both the optical/near-infrared (IR) and (sub-)millimeter regimes. While sub-mm and mm observations mostly probe the disc mid-plane, near-IR observations trace the surface layers of the disc. 

The focus of this paper is the protoplanetary disc around the T Tauri star \hd{}. The central object is a G-type star, with temperature $T_{\rm eff}$~=~5880~K, luminosity $L_{*}$~=~4.58~L$_{\odot}$ and mass $M_{*}$~=~1.8$^{+0.2}_{-0.3}$~M$_{\odot}$ \citep{2018A&A...619A.171B, 2018ApJ...869L..41A}. It is located at a distance of 165~$\pm$~5~pc \citep{2018A&A...616A...1G}, in the region of Upper Sco, and it is relatively old \citep[4-12 Myr,][]{2012ApJ...746..154P}. Non-detections in interferometric \citep{2008ApJ...679..762K} and high-contrast imaging \citep{2018A&A...619A.171B} observations set an upper limit on the mass of the companion to be $\sim 0.4\,M_{\odot}$. This corresponds to a limit on the mass ratio of any putative binary of $q\lesssim 0.2$. 
Any objects more massive than this should have been detected in the IR. Additionally, disc structures in the polarized scattered light image are resolved down to an angular separation of $\sim$~0.06''. For any binary separation smaller than $\sim$~10~au, it would not be possible to distinguish the two stars.

The disc around \hd{} is seen close to face-on, and has recently been studied in detail at multiple wavelengths. \cite{2018A&A...619A.171B} presented polarized scattered light $J$-band images observed with VLT/SPHERE, with an angular resolution of $\sim$0.037" (6.1~au). \hd{} has also been observed at very high resolution (0.046", i.e. 7.6~au) as part of the DSHARP survey \citep{2018ApJ...869L..41A} and its 1.25~mm continuum emission has been analyzed by \cite{2018ApJ...869L..50P}. 

The origin of most of the structures visible in the disc are not yet fully understood. Here we briefly summarize the main detected morphological features:
\begin{enumerate}[label=(\alph*), left=0pt .. \parindent]
    \item The IR scattered light image shows an  asymmetric ring-like feature between 18 and 30~au that presents two narrow shadows along the north-south direction. A broader dark region, covering the western half of the disc, can also be seen \citep{2018A&A...619A.171B}. \label{itm:shadows}
    \item The SPHERE image also shows an over-brightness at about PA~$\sim$150$^{\circ}$ \citep{2018A&A...619A.171B}. \label{itm:arcsphere}
    \item The scattered light observations imply a relative misalignment of $\sim$~30$^{\circ}$ between the inner and outer discs \citep{2018A&A...619A.171B}.\label{itm:relmis}
    \item The 1.3~mm continuum emission from the disc resolves into three bright rings at roughly 8, 40 and 64~au from the disc center, and a large gap between $\sim$10 and 30~au \citep{2018ApJ...869L..50P, 2018ApJ...869L..41A}. \label{itm:gap}
    \item A bright arc in the south-east is also visible in the dust observations, outside of the outermost ring at $\sim$74~au \citep{2018ApJ...869L..50P, 2018ApJ...869L..41A}. \label{itm:arcalma}
    \item Kinematic observations of the CO line reveal a ``kink'' in the red-shifted channel that is not present in the blue-shifted channel \citep{2018ApJ...869L..50P, 2018ApJ...869L..41A, 2020ApJ...890L...9P}. \label{itm:kink}
\end{enumerate}

Most of these sub-structures suggest that the inner disc is misaligned, which could be due to either a central binary or a planet located further out in the disc.
Previous 3D hydrodynamic simulations have already shown that an inclined binary causes disc breaking \citep[e.g.][]{2013MNRAS.434.1946N, 2013MNRAS.433.2142F}, producing azimuthal asymmetries in scattered light observations \citep[e.g.][]{2018MNRAS.473.4459F, 2018MNRAS.475L.35M, 2020MNRAS.493L.143N}. Indeed, the disc breaks into two distinct components, with the resulting inner disc misaligned with respect to the outer regions. The inner disc precesses as a rigid body while the outer disc evolves similarly on a longer time-scale. Such a configuration produces a unique signature in IR observations. The light of the star is blocked by the inner disc, which is optically thick to such wavelengths and casts a shadow on to the outer disc. The morphology of the SPHERE image of \hd{} was approximately recovered using 3D hydrodynamic simulations of a disc warped by a misaligned equal mass binary \citep{2018MNRAS.473.4459F, 2018A&A...619A.171B}, though such a massive companion is ruled out by existing IR observations \citep{2008ApJ...679..762K, 2018A&A...619A.171B}.

Alternatively, a misaligned planetary companion has been suggested to help explain the disc structure. Such a planet would have a mass equal to 10-100~M$_{\rm Jup}$ (i.e. mass ratio 0.01-0.1), potentially located between 8 and 40~au. 
As suggested by \cite{2018A&A...619A.171B}, for example, a 10 Jupiter-mass planet around a 1.5~M$_{\odot}$ star would not be detected by interferometric observations, and at the same time, may cause the misalignment of an inner disc. 
The hypothesis of a planet hosting disc is strongly suggested by the ALMA dust continuum image, which shows a large gap at about $\sim$22~au. As already inferred by \cite{2018ApJ...869L..47Z}, an embedded object with a mass of about $\sim$10-20~M$_{\rm Jup}$ could easily produce a dust depletion as expected in the sub-mm observations. Moreover, \cite{2018ApJ...869L..50P} analyzed the CO emission and found a deviation from the Keplerian rotation pattern in the red-shifted channels, at a radial distance of about 32~au. This, again, hints at the presence of a planetary companion, within the annular dust gap. \citet{2019NatAs...3.1109P} showed that embedded planets perturb the disc gas, resulting in a kinematic signature which is detectable with high resolution observations. They found that the minimum detectable mass in kinematic observations is roughly 2~M$_{\rm Jup}$. In a kinematic study of multiple sources within the DSHARP sample, \citet{2020ApJ...890L...9P} found that \hd{} has significantly larger velocity perturbations than the other sources, which points towards the presence of a planet more massive than 1-3~M$_{\rm Jup}$. 

The combination of the previous configurations has never been considered so far. If existent, it would imply a planet orbiting a binary. In this scenario, planets are generally classified as S-type and P-type planets depending on the perturbation induced by the host stars. S-type (or circumstellar) planets orbit around one of the stars of the system, while P-type (or circumbinary) planets orbit around both stars. The occurrence rate of the two populations is very different, where more than 120 are known as circumstellar planets while only a dozens are circumbinary planets \citep{2019Galax...7...84M}. 
Additionally, planets around binary stars may have their orbits inclined with respect to that of the binary. This explanation has been invoked to interpret systems like 16~Cygn~B. The eccentric planet was initially thought to be on a circular inclined orbit, which then became aligned and eccentric due to Kozai Lidov oscillations \citep{1997Natur.386..254H}. Systematic studies looking at the stability of S-type planets are being developed, considering the properties of the binary as well as the planet distance and inclination \citep[e.g.,][]{2020AJ....159...80Q}. On the contrary, similar studies for P-type planets have not been exploited so far.

We will show in this paper that a single companion, of either stellar or planetary mass, cannot account for the bright arc seen in both the SPHERE \ref{itm:arcsphere} and ALMA \ref{itm:arcalma} observations at about the same azimuthal location, but different radii. In this paper we reconsider both of these scenarios, and their limitations, and instead propose that the \hd{} disc hosts both a misaligned stellar binary and a giant circumbinary planet. 

\section{Numerical methods}
\label{sec:methods}
We perform 3D hydrodynamic simulations using the SPH code \phantomcode{} \citep{2018PASA...35...31P}. We perform a suite of numerical simulations to investigate scenarios to explain the morphological features observed in the disc surrounding \hd{}. 
Common to all our simulations, we model a gaseous disc of mass 0.01~M$_{\odot}$ and an initial radial extension from 4~au out to 100~au. The gas surface density profile follows $\Sigma \propto r^{-1}$, and we adopt a locally isothermal equation of state, with a temperature profile given by $T\propto r^{-0.5}$. The disc aspect ratio is initially set to $H/R=0.057 (R/R_{\rm ref})^{0.25}$, where the reference radius is R$_{\rm ref}$~=~30~au. 
We follow the \cite{1973A&A....24..337S} prescription for $\alpha$ by fixing $\alpha_{\rm SS}$~=~0.005 at our chosen resolution, and with a corresponding alpha viscosity parameter $\alpha_{\rm av} \approx~0.2$ \citep{2010MNRAS.405.1212L}.

For Section~\ref{sec:case3} we also include a population of 1~mm dust grains. We compute the dust dynamics using the one-fluid algorithm \citep{2014MNRAS.440.2136L,2015MNRAS.451..813P,2018MNRAS.477.2766B}. We initially set the dust radial density profile equal to the gas profile, with a gas-to-dust ratio of 100. 

\begin{figure*}
	\centering
	\includegraphics[width=\textwidth,trim={0cm 0cm 0cm 0cm},clip]{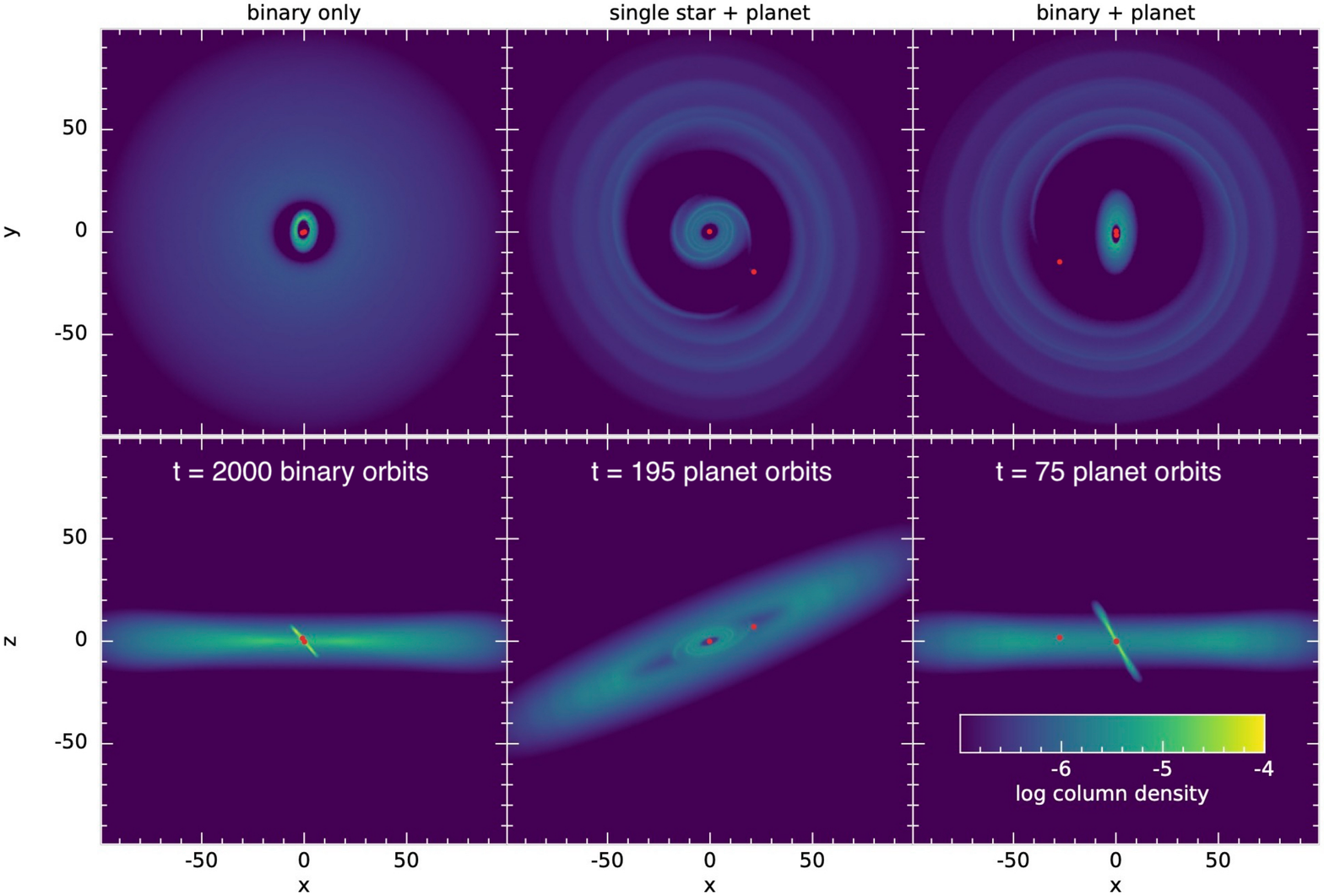}
    \caption{Snapshots of the hydrodynamic simulations for the three scenarios presented in Section~\ref{sec:results}: a binary inclined by 60$^{\circ}$, with $q=0.2$ and $a=2$~au (first column), a 20 Jupiter-mass planet located at R$_{\rm p}$~=~32~au around a single star and misaligned by 30$^{\circ}$ (second column) and a planet orbiting an inclined binary, where we combine the properties of the previous configurations (third column). The first and second rows show a face-on and an edge-on view of the gas density distribution, respectively. The sink particles are indicated by the red dots. We have chosen a representative physical time for each situation.
	}
    \label{fig:scenarios}
\end{figure*}

In order to compare our model with observations of \hd{}, we post-process our simulations using the Monte Carlo radiative transfer code MCFOST \citep{2006A&A...459..797P,2009A&A...498..967P}. The code uses a Voronoi mesh where each cell corresponds to the position of an SPH particle. We use $10^8$ photon packets to compute the dust temperature. We attribute the known stellar parameters to the primary star and calculate the radius from temperature and luminosity. We then assume a 5~Myr isochrone \citep{2000A&A...358..593S} and derive the properties of the secondary star. The masses and positions are updated directly from the simulation. Figure~\ref{fig:temp_midplane} in Appendix~\ref{tempappendix} shows a comparison between the temperature computed by MCFOST and assumed in \phantomcode.

We compute synthetic CO channel maps of the $^{12}$CO $J$=2--1 line (assuming  a CO-to-H$_2$ abundance of 1$\times 10^{-4}$, T$_{gas}$ = T$_{dust}$, and that the population levels are at LTE). The synthetic CO maps have a spectral resolution of $\sim$100~m/s.
We also compute the corresponding simulated IR scattered light and 1.3~mm continuum images. For this calculation, we consider a population of grains with size ranging from 0.03 to 1000~$\mu$m. Dust grains smaller than 1$\,\mu$m are assumed to follow the gas distribution, and in each cell we compute the density for each grain size by interpolating the SPH densities between 1$\,\mu$m and 1mm. The grain size distribution over the whole disc is then normalized following a power law d$n(a) \propto a^{-3.5}$ d$a$, and a dust-to-gas ratio of 100.
The dust optical properties are determined using the Mie theory. Following \cite{2018A&A...619A.171B}, we convolve the IR image with an angular resolution of $\sim$~0.037".
We then perform synthetic observations of the sub-mm flux, using the software CASA \citep{2007ASPC..376..127M}. We use an integration time of 35 minutes and a beam size of $\sim$52~$\times$~45~mas consistent with the observations, using the corresponding preset configuration. We measure an RMS noise of $\sim$6.2~10$^{-2}$ mJy beam$^{-1}$.

\section{Models}
\label{sec:results}
We investigate three different scenarios in turn, as motivated by the features observed in \hd{}: i) an inclined inner binary, ii) a misaligned planet around a single star and iii) an inclined binary with a co-planar planet. A snapshot of a simulation for each scenario is shown in Figure~\ref{fig:scenarios}. The first, second and third columns correspond to the cases presented in Sections~\ref{sec:case1}, \ref{sec:case2} and \ref{sec:case3}, respectively. 
The success of each scenario is measured by comparison with the observed features. We present the density evolution and synthetic observations only for our final model.

\subsection{An inclined binary} \label{sec:case1}
We first test the possibility that the misalignment of the central region is driven by an inclined binary system, as suggested in \cite{2018A&A...619A.171B}, based on \cite{2018MNRAS.473.4459F}. 

\begin{figure*}
	\centering
	\includegraphics[width=\textwidth,trim={3.3cm 0cm 4.cm 1.2cm},clip]{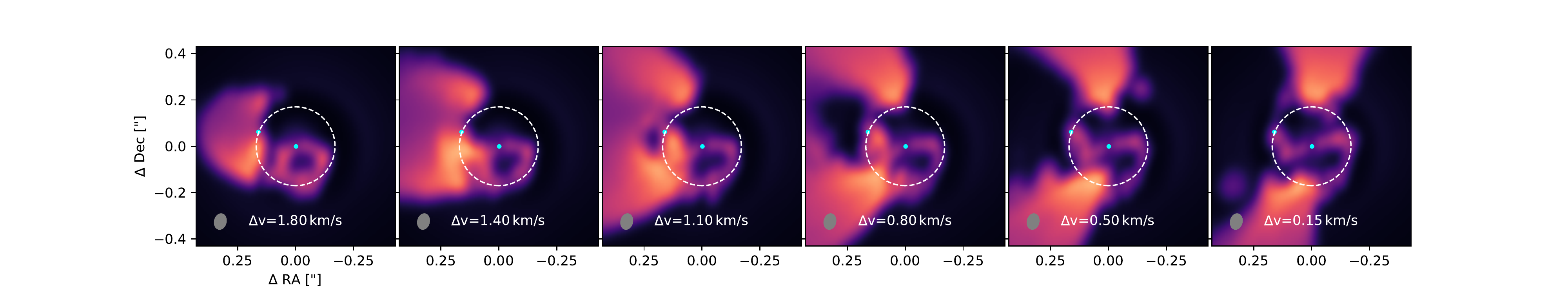}
	\caption{Red-shifted channel maps of the $^{12}$CO $J$=2--1 emission for the ``single star+planet'' model, calculated subtracting the systemic velocity. Here we show the results from the modelling presented in Section~\ref{sec:case2}. The cyan dots indicate the position of the central star and the 20~M$_{\rm Jup}$ planet, inclined by 30$^{\circ}$ to the outer disc. The gas velocity pattern is perturbed by the misaligned material interior to the planet's orbit (indicated by the dashed white circle). The beam size is represented by the grey ellipse in the bottom left corner of each channel.}
	\label{fig:channel_maps_inclined}
\end{figure*}

Taking into consideration the detection limits of near-IR observations (see Section~\ref{sec:intro}), we perform a set of hydrodynamic simulations with a gaseous disc and a central binary of mass ratio $q$~=~0.1, 0.2, 0.3 and separation $a$~=~1, 2 and 5~au. The binary is initialised on a circular orbit, inclined by 60$^{\circ}$ to the disc plane. Here, we deliberately choose a large inclination for the binary as \citet{2018MNRAS.473.4459F} showed that this larger inclination guarantees a wide range of relative misalignments, including the 30-40$^{\circ}$ suggested by the observations as \cite{2018A&A...619A.171B}. We let the system evolve for a few thousand binary orbits and consider the resulting disc structure. 

As predicted with Eq.~9 of \cite{2013MNRAS.434.1946N}, we find that the disc breaks at the expected radius in our simulations with $a$~=~2 or 5~au and $q$~=~0.2 or 0.3. In particular, in the simulation with $a$~=~2~au and $q$~=~0.2 the disc breaks at roughly 8~au, which corresponds to the upper limit in the observations. A snapshot from this simulation is shown in the left panels of Figure~\ref{fig:scenarios}, after about 2000 binary orbits. 
None of these simulations produce the extended dust gap nor explain the kink in the channel map found around 32~au. We thus conclude that a misaligned binary is not sufficient to explain the observed morphology of \hd{}.

\subsection{A misaligned planet around a single star} \label{sec:case2}
Our second scenario considers an inclined companion orbiting around a single star that can tilt the inner disc, casting a shadow onto the outer disc. We therefore model a disc around a single star of mass 1.8~M$_\odot$ and a misaligned planet at R$_{\rm p}$~=~32~au. We consider inclinations of 15 and 30 degrees and four planetary masses, M$_{\rm p}$~=~2, 5, 10 and 20~M$_{\rm Jup}$. We initialise the disc with a gap of width $\Delta r$~=~2~au, centered around the location of the planet, to prevent unphysically rapid gas accretion at the beginning of the simulation. For this model, we evolved the hydrodynamic simulations for $\sim$300 orbits of the planet and then we computed the CO channel maps. All the considered planets are massive enough to leave a signature above the detection limit of the kinematic observations.  

Looking at the results of our hydrodynamic simulations, we find that planets with masses of 2 and 5~M$_{\rm Jup}$ produce a maximum relative misalignment between inner and outer disc of 4$^{\circ}$, while the 10~M$_{\rm Jup}$ planet reaches a relative misalignment of about 12$^{\circ}$. Only the 20 Jupiter-mass planet on a 30$^{\circ}$ inclined orbit produces the large relative misalignment between inner and outer discs required by the scattered light observations. Indeed, after about 100~orbits of the planet, we measure a relative misalignment of 30$^{\circ}$. However, the planet's angular momentum exceeds the angular momentum of both the inner and outer disc. As a result, the outer disc also moves in response to the planet, tilting away from its initial position and aligning with the rest of the disc. The middle panels of Figure~\ref{fig:scenarios} show a snapshot of the 20 Jupiter-mass planet inclined by 30 degrees after orbiting $\sim 195$ orbits. Thus a planet massive enough to make the required misalignment between the inner and outer disc is also massive enough to tilt the outer disc, and as such the misalignment is not sustainable on time-scales greater than about 200 planet orbits. 

In all of our misaligned planet simulations that develop an inclined inner disc, because the inner and outer disc are separated at the orbit of the planet all of the material interior to the planet orbit becomes misaligned. 
This results in a velocity perturbation which is visible in the simulated channel maps at all radii r~$\leq$~R$_{\rm p}$. An example is illustrated in Figure~\ref{fig:channel_maps_inclined}. 
However, Fig.~7 in \cite{2018ApJ...869L..50P} shows only a kink in the red-shifted channel, at the azimuthal and radial position of the planet. No other deviations or evidence of misalignment from the Keplerian pattern are visible. While a misaligned planet may lead to a misaligned inner disc, the suggested location of the planet and indication of where the disc is misaligned are not consistent with this idea. We therefore conclude that a misaligned planet only is also not able to explain all of the features in \hd{}.

\subsection{An inclined binary with a planetary companion} \label{sec:case3}
As the previous two scenarios cannot explain the data, we propose a third option: an inclined binary with an outer planetary companion. As mentioned in Section~\ref{sec:case1}, an inclined binary can warp the disc, breaking and driving precession of the inner disc. The planet instead is able to open a gap in the outer region of the disc. We thus seek to combine elements of these previous scenarios to simultaneously explain all the features identified in \hd{}. For this calculation, we select the most promising parameters from previous simulations. We adopted a binary separation of 2~au, a mass ratio 0.2 and an orbital inclination of 60$^{\circ}$. We embedded a planet at 32~au with mass 10~M$_{\rm Jup}$, on an orbit co-planar with the outer disc. For this scenario only we additionally included a dust component, described in Section~\ref{sec:methods}. 

\begin{figure*}
	\centering
	\includegraphics[width=\textwidth,trim={0cm 0cm 0cm 0cm},clip]{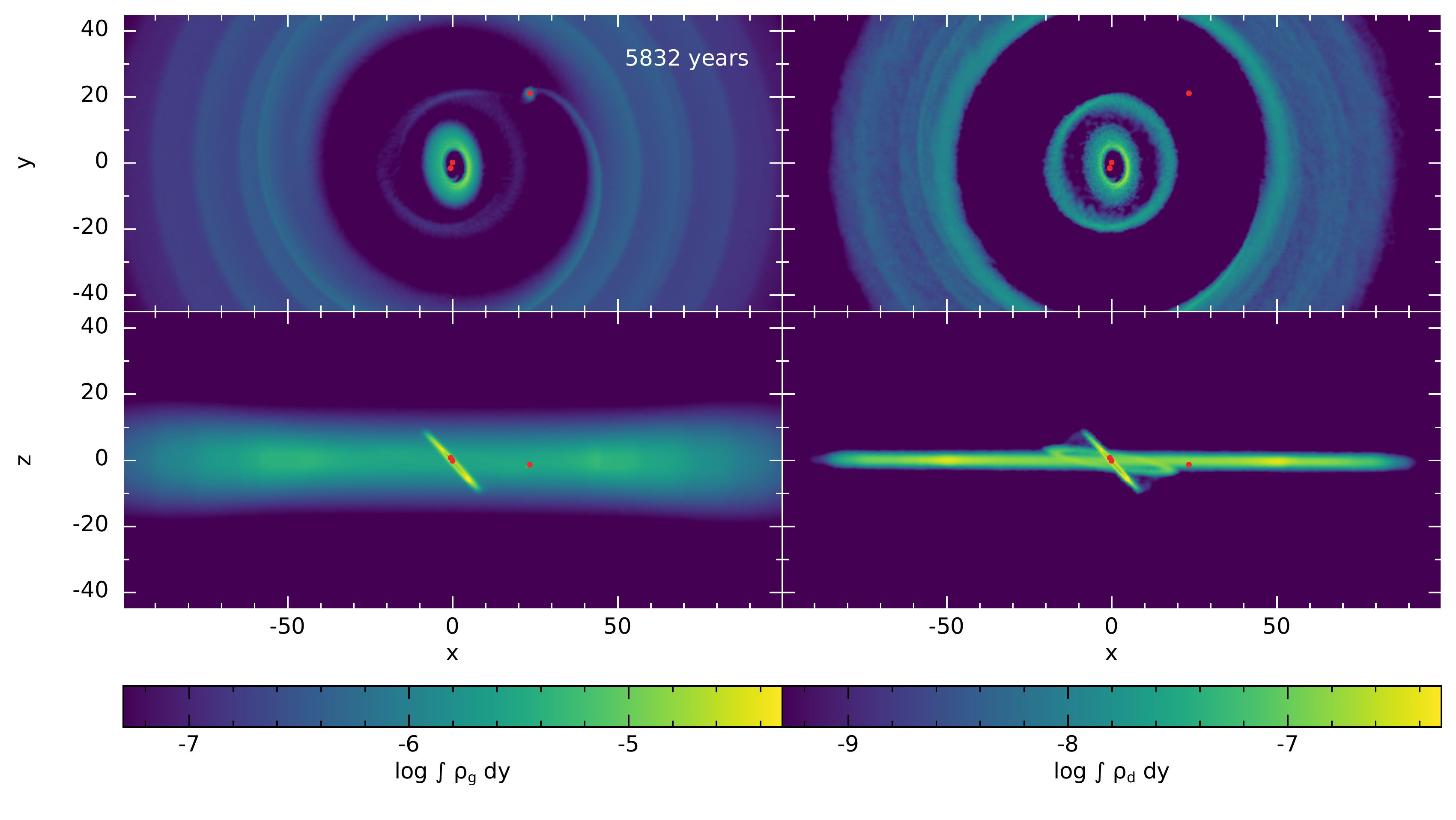}
	\caption{Snapshots of the ``binary+planet'' hydrodynamic simulations, showing a face-on (upper panels) and an edge-on view (lower panels) of a planet hosting disc, with a misaligned central binary. We present the gas and dust surface density profiles in the left and right columns, respectively. The red dots represent the sink particles in our simulation.}
	\label{fig:density_render}
\end{figure*}

Figure~\ref{fig:density_render} shows a face-on (upper panel) and edge-on (lower panel) density rendered view of this planet hosting disc with a central misaligned binary. 
As in Section~\ref{sec:case1}, we find that after a few hundred years, the inner disc breaks and start precessing away from the outer disc. 

We derive the precession time by measuring how long it takes for the disc to twist through a full $2\pi$ radians. Following \citet[][Eq.~A2 in the Appendix]{2015MNRAS.448.1526N}, we measure the gradient of the twist and find that the precession time-scale increases from 400~yr at 1000~yr to 1300~yr at 3500~yr. The precession slows down as the inner disc tends to align with the binary orbit. 
We find a similar result when using Eq.~4 in \cite{2018MNRAS.473.4459F}. This relationship is very sensitive to the inner edge of the disc, and for an inner radius of 4~au we derive a precession time-scale of $\sim$~1100~yr.
Likewise, we calculate the precession time of the outer disc, which is much slower, and find that it is $\sim 1.7 \times 10^5$~years.

After approximately 50 planetary orbits, an annular dust gap has formed within the disc, extending out to 40~au. The disc remains stable for about hundred orbits of the planet. Considering that circumbinary orbits are generally stable especially far from the center and the planet is not  massive, we expect the system to survive in this configuration also on a longer timescale. 
\cite{1999AJ....117..621H} have investigated the long term stability of planets in binary systems. In particular, they derived the critical semi-major axis beyond which orbits are dynamically stable. Interestingly, when considering P-type planets, non co-planar orbits, such as polar, are stable over a wide range of binary properties \citep{2019A&A...628A.119C, 2020MNRAS.494.4645C}. 
We note, however, that if the planet is massive enough and close to the central star, it can cause the binary to precess \citep[e.g.,][]{2019A&A...628A.119C, 2019MNRAS.490.5634C}.
During the simulation, a fraction of the disc mass is also accreted onto the planet, forming a circumplanetary disc, misaligned to the orbital plane.

\subsection{Synthetic observations}

\subsubsection{``Choosing'' the observed configuration}
\begin{figure}
	\centering
	\includegraphics[width=0.47\textwidth,trim={0cm 0cm 0cm 0cm},clip]{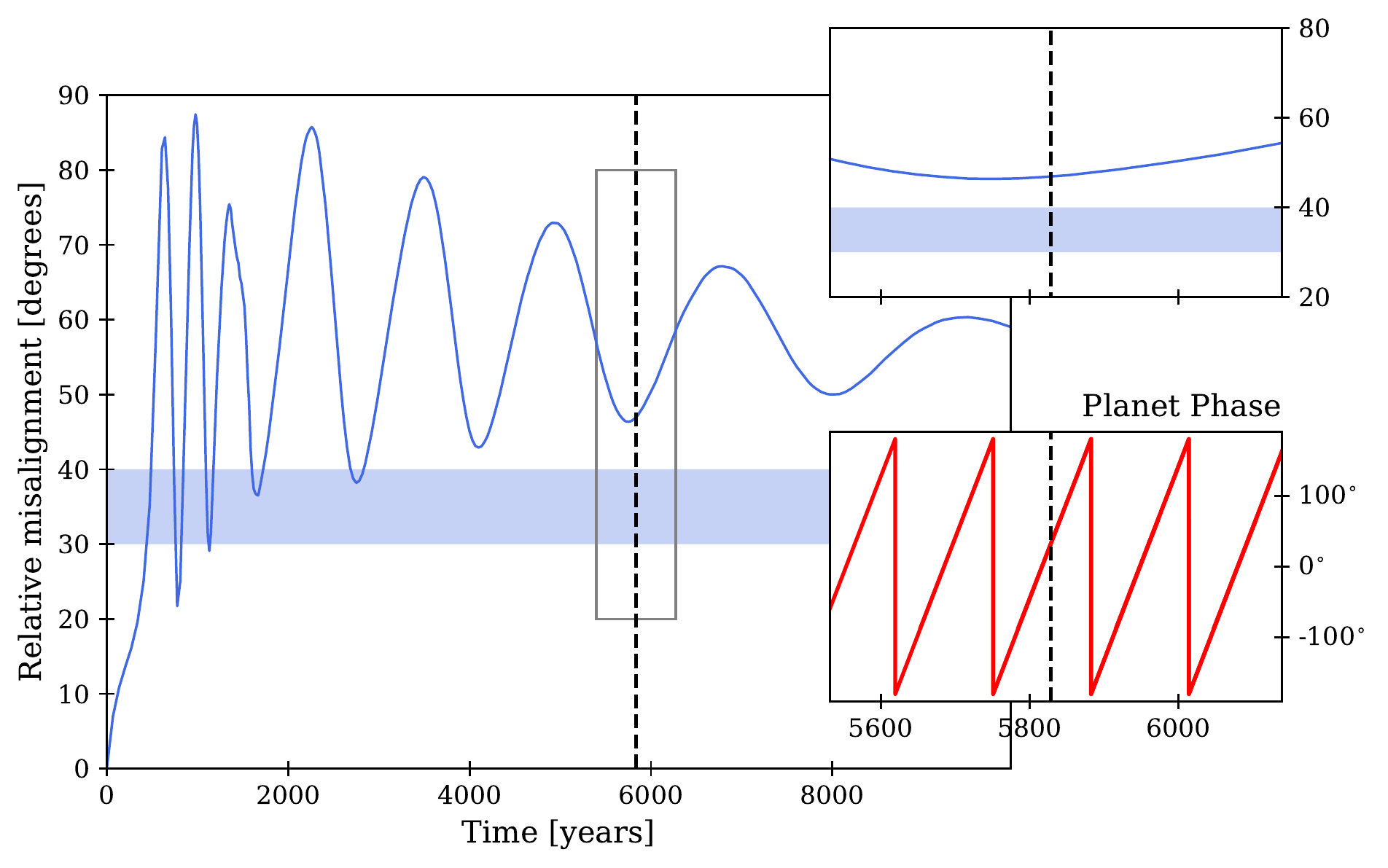}
	\caption{Time evolution of the relative misalignment between inner and outer discs. The shaded blue area represents the range of relative inclinations inferred from observations of HD143006 \citep{2018A&A...619A.171B}. The upper inset is a zoom in of the relative misalignment in the time period highlighted by the gray rectangle. The lower inset shows the planet's orbital phase (expressed in degrees)  over the same time period. The dashed black vertical line at $t=5832$~yr indicates the time at which we select the snapshot shown in Figure~\ref{fig:density_render}.}
	\label{fig:relative_misal}
\end{figure}

\begin{figure*}
	\centering
	\begin{subfigure}
	    \centering
	    \includegraphics[width=0.47\textwidth,trim={0cm 6cm 0cm 6cm},clip]{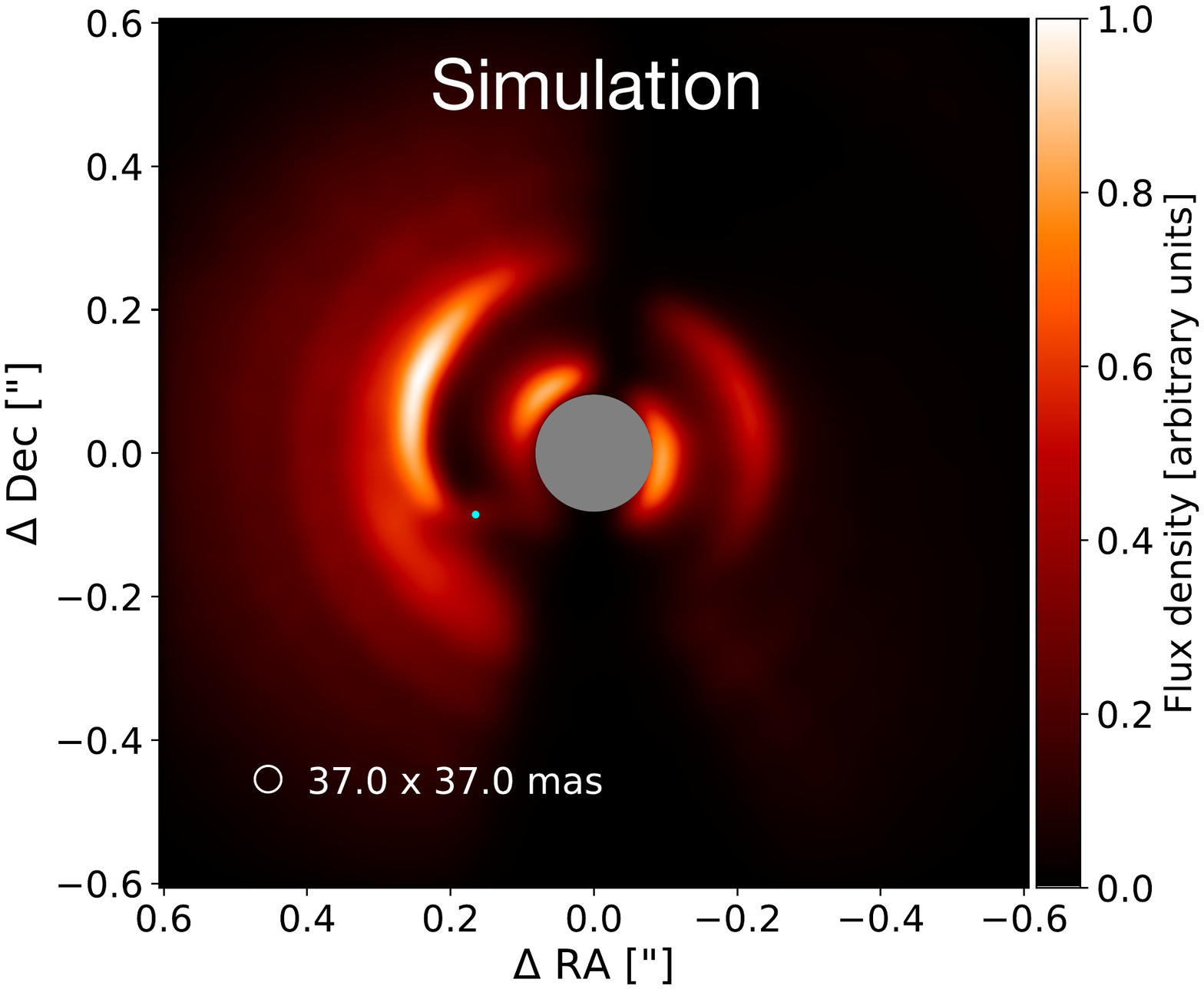}
	    \includegraphics[width=0.48\textwidth,trim={0cm 6cm 0cm 6cm},clip]{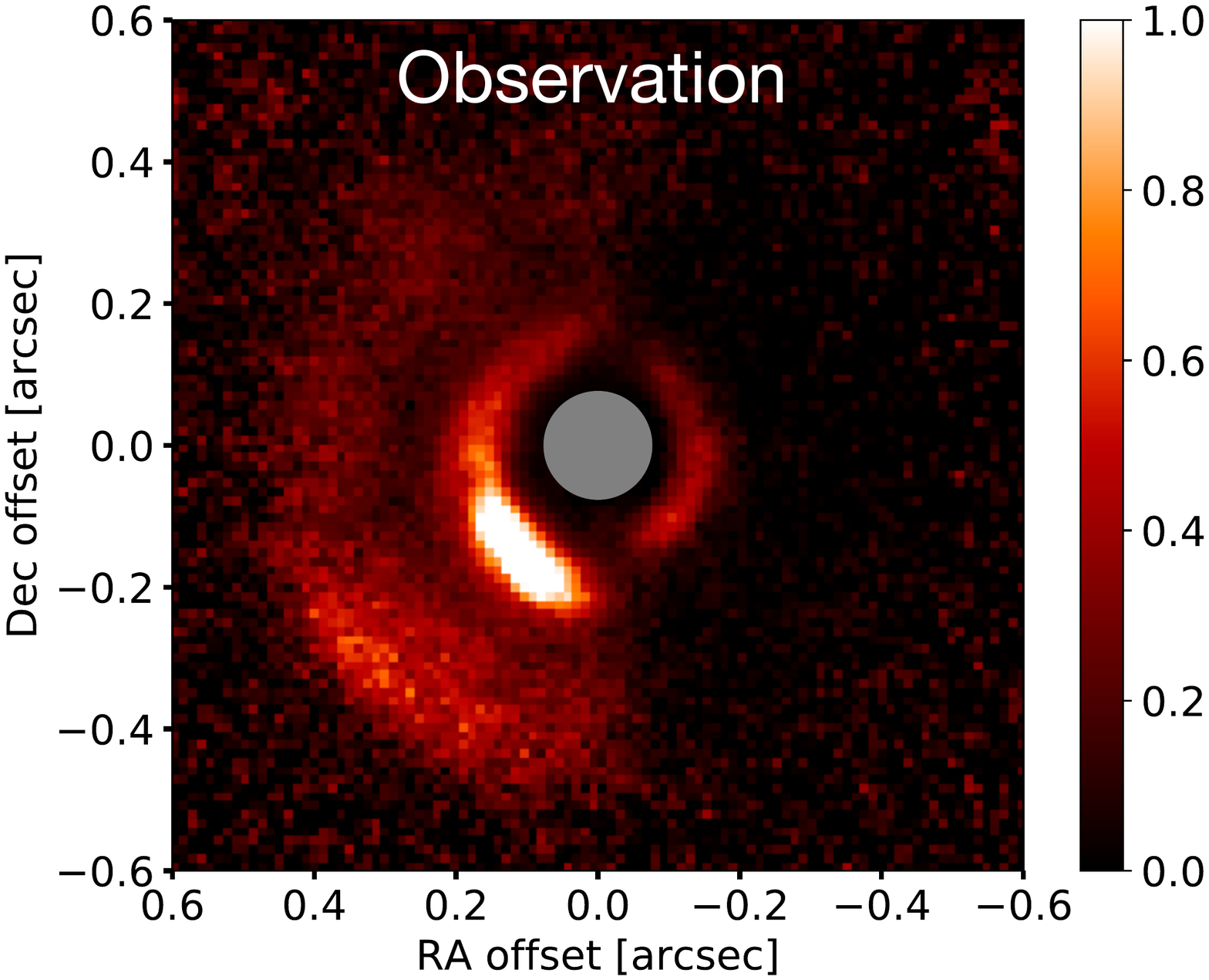}
	    \includegraphics[width=0.50\textwidth,trim={0cm 6cm 0.5cm 7cm},clip]{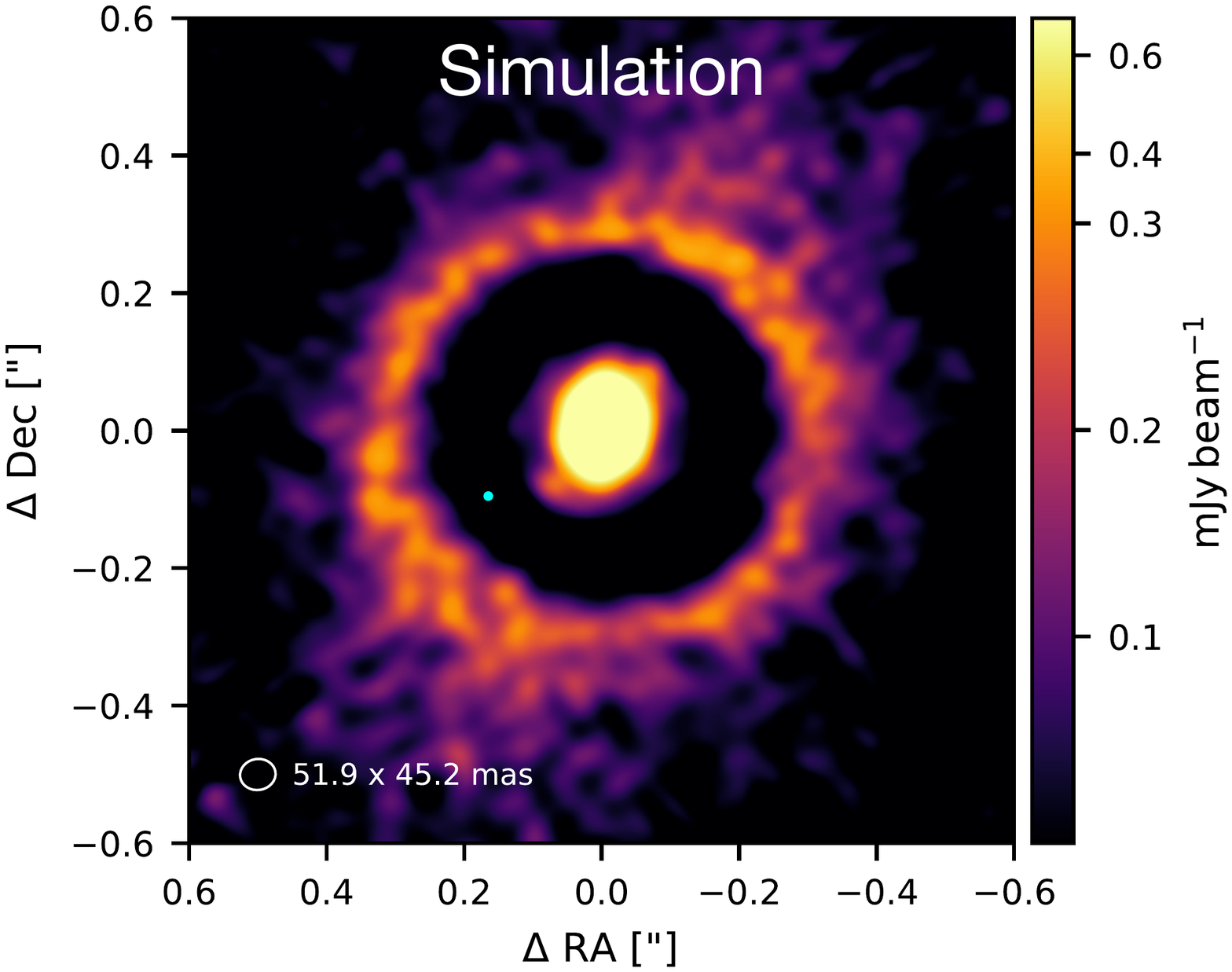}
	    \includegraphics[width=0.48\textwidth,trim={0.1cm 8cm 6cm 9cm},clip]{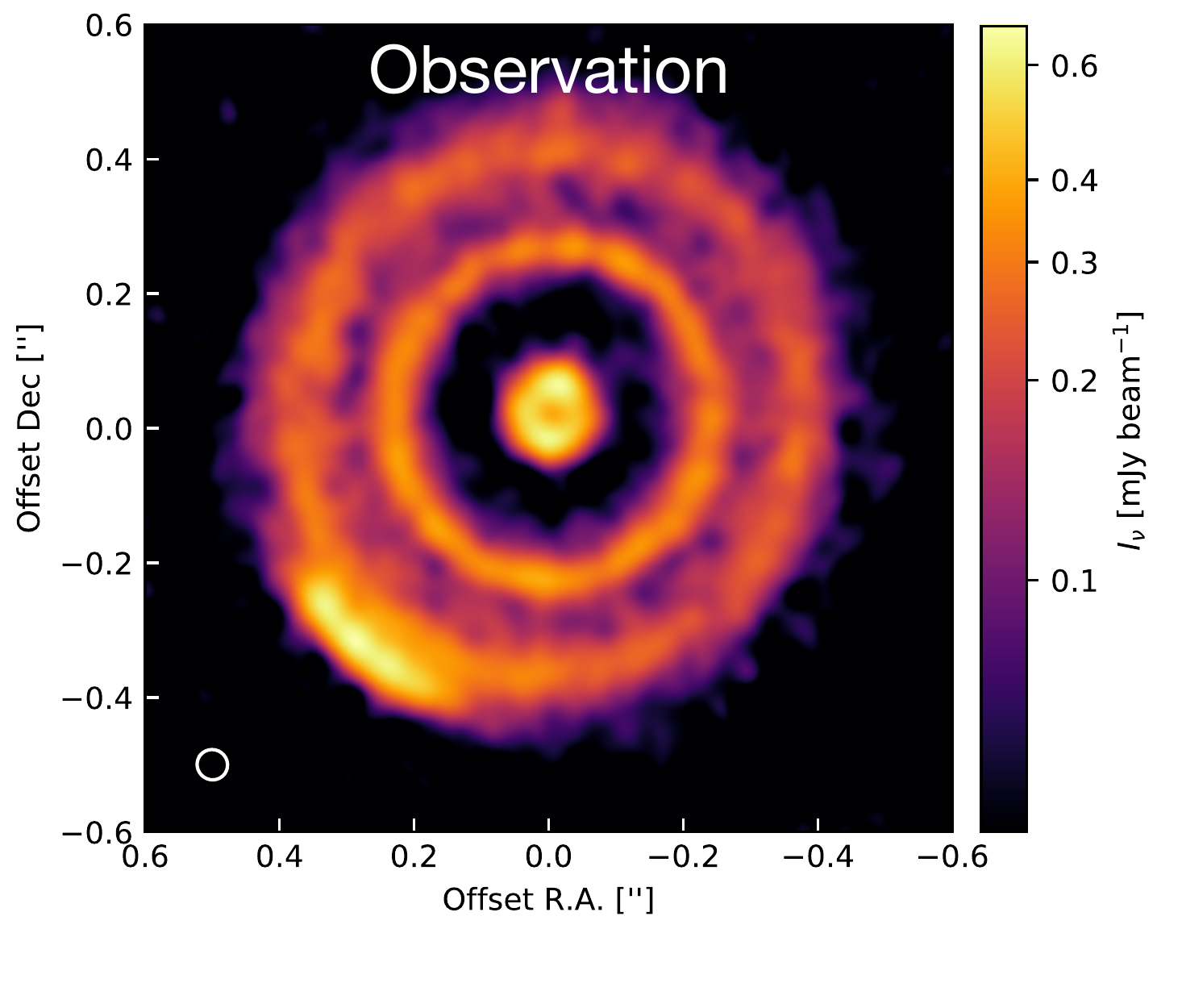}
	    \caption{Comparison between the results from our ``binary+planet'' model (left panels) and the observations (right panels). Top panels: Polarized intensity image $Q_{\Phi}$ observed at 1.2~$\mu$m ($J$-band) of the hydro simulation of \hd{}. The flux has been scaled by $r^2$ to enhance the drop in intensity. The gray circle indicates the 0.08" coronographic mask. The image on the right has been taken from \protect\cite{2018A&A...619A.171B} \protect\footnotemark[1]. Bottom panels: ALMA Band 6 synthetic observation at 1.3~mm of the dust continuum emission. The empty circles in the bottom left corners indicate the corresponding beam size. The image on the right has been taken from \protect\cite{2018ApJ...869L..50P} \protect\footnotemark[2].}
    	\label{fig:sphere_alma}
	\end{subfigure}
\end{figure*}

Observable features such as shadows and kinks depend on the relative phases of inner/outer disc precession, and also on the phases of the planet and binary orbits. We therefore only expect to reproduce these observations at particular times in our simulation, when all these phases are similar to the current configuration of the HD143006 system. There are effectively two angles which determine the key observables: the relative misalignment between the inner and outer disc; and the orbital phase of the planet. The orientation of the inner disc primarily affects the IR scattered light and mm continuum emission, while the position of the planet determines the location of the kink in the CO channel maps.

The evolution of the relative misalignment between the inner and outer discs is shown in Fig.~\ref{fig:relative_misal}. Once it breaks, the inner disc completes several full precessions during the course of the simulation, and the relative misalignment is gradually damped as the inner disc aligns with the binary (which, as discussed in Section \ref{sec:case1}, was initially inclined at 60 degrees). 
If the binary is eccentric and misaligned, the inner disc could become polar \citep{2015MNRAS.449...65A, 2017ApJ...835L..28M, 2018MNRAS.479.1297M, 2018ApJ...869L..47Z, 2019A&A...628A.119C}. If this happens, then the mutual misalignment between the three components is long-lived.   
The best match with the observed misalignment is therefore found close to the minimum of each precession cycle. As we have not fine-tuned the simulation parameters there is no ``perfect'' snapshot which reproduces all of the observations precisely, but we see closest agreement at several different times during the simulation.  Our model also relies on the dust clearing between $8-30$~au by the planet. This gap has mostly been cleared by $t\sim7000$~yr, so we restrict our choice to later times in the simulation. We therefore choose to show results from the configuration at $t=5832$~yr (shown in Fig.~\ref{fig:density_render}): this time represents the ``best compromise'' in matching the observations. It corresponds to 43 orbits of the planet and 2766 binary orbits. 
At this time the inner disc has a relative inclination of $\simeq 47$ degrees with respect to the outer disc, which is consistent with the IR scattered light observations \citep{2018A&A...619A.171B}, and the continuum emission and CO velocity field \citep{2018ApJ...869L..50P}. Additionally, the planet is located at the approximate location of the kink suggested by \citet{2018ApJ...869L..50P} and confirmed by \citet{2020ApJ...890L...9P}. Using this configuration of the system, as shown in Fig.~\ref{fig:density_render}, we now consider the different synthetic observations in turn.

\begin{figure*}
	\centering
    \includegraphics[width=\textwidth,trim={3.cm 0.2cm 2.cm 0.2cm},clip]{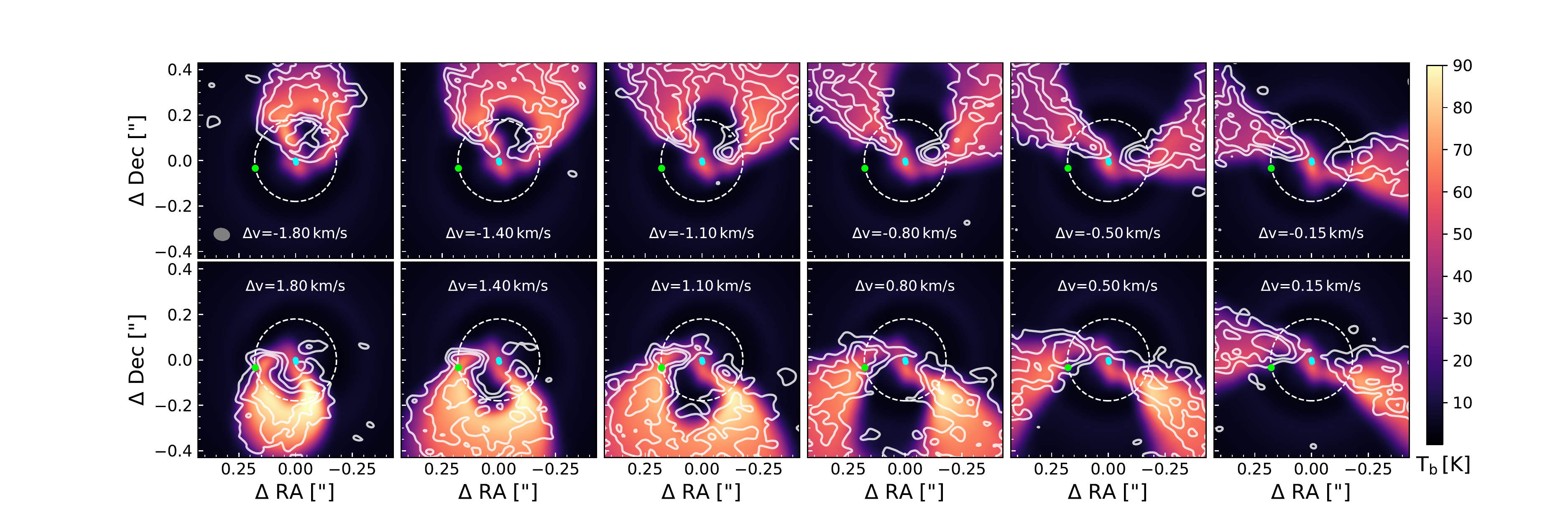}
	\caption{Synthetic channel maps of the $^{12}$CO $J$=2--1 emission, calculated subtracting the systemic velocity. Top and bottom rows are the blue and red-shifted channels respectively. The overlaid white contours are the kinematic observations as presented in Fig.~7 of \protect\cite{2018ApJ...869L..50P}. The cyan dots indicate the positions of the central binary, while the green dot and the white dashed circle indicate the position and orbit of the planet. The beam resolution of 66 $\times$ 49~mas is represented by a grey ellipse in the top left panel. The spectral resolution is 100 m s$^{-1}$.
	}
    \label{fig:kinematics}
\end{figure*}

\footnotetext[1]{Credit: \cite{2018A&A...619A.171B}, reproduced with permission \textcopyright~ESO.}
\footnotetext[2]{Credit: \cite{2018ApJ...869L..50P}, reproduced with permission \textcopyright~AAS.}

\subsubsection{Infrared emission} \label{sec:sphereobs}
We computed synthetic observations from our hydrodynamic simulation of \hd{} with both the binary and the planet. 
In the upper left panel of Figure~\ref{fig:sphere_alma} we present a synthetic scattered light observation at 1.2~$\mu$m ($J$-band). Following \citet{2018A&A...619A.171B}, the flux has been scaled by $r^2$ to enhance the drop in intensity, along with a coronographic mask. 
The image shows a left/right brightness asymmetry, along with a bright arc due to the illumination of the inner edge of the outer disc. The misaligned inner disc obscures the light from the central binary on one side of the disc, causing the wide shadow across half the disc. This is accompanied by two narrow lane shadows cast by the mid-plane of the inner disc. The two smaller arcs closest to the coronographic mask are caused by an illumination effect of the material around the inner disc, which has not fully accreted on to the inner disc (as shown in Figure~\ref{fig:density_render}). Our simulations show that this feature disappears after $\sim$~7000 yr, once the dust has accreted onto the central disc.

\subsubsection{CO kinematics} \label{kinematics}
Figure~\ref{fig:kinematics} shows our synthetic velocity channel maps for the CO emission. The overlaid white contours are the kinematic detections as presented in Figure~7 of \cite{2018ApJ...869L..50P}. The velocity in each channel is calculated relative to the systemic velocity. Top and bottom rows show the blue and red-shifted channels, respectively. A kink is visible in some of the red-shifted channels, and coincides with the radial and azimuthal position of the planet (indicated in cyan).
This observational signature has been previously attributed to the presence of a Jupiter-mass planet within the disc \citep[e.g.][]{2019NatAs...3.1109P, 2020ApJ...890L...9P}, and it is confirmed by the model presented here. 
We also notice an asymmetry in the wings of the butterfly-shaped flux distribution, which can also be traced in the observations in Figure 7 of \cite{2018ApJ...869L..50P}. 
In this context, the stellar companion could also be responsible for a perturbation of the velocity pattern. These effects could not be easy to disentangle from other perturbations, especially considering the lower resolution of the CO observations. Nonetheless, we expect these effects to be very close to the central binary.

\subsubsection{Sub-mm dust emission} \label{almaobs}
\begin{figure}
	\centering
	\includegraphics[width=\columnwidth,trim={0cm 0cm 0cm 0cm},clip]{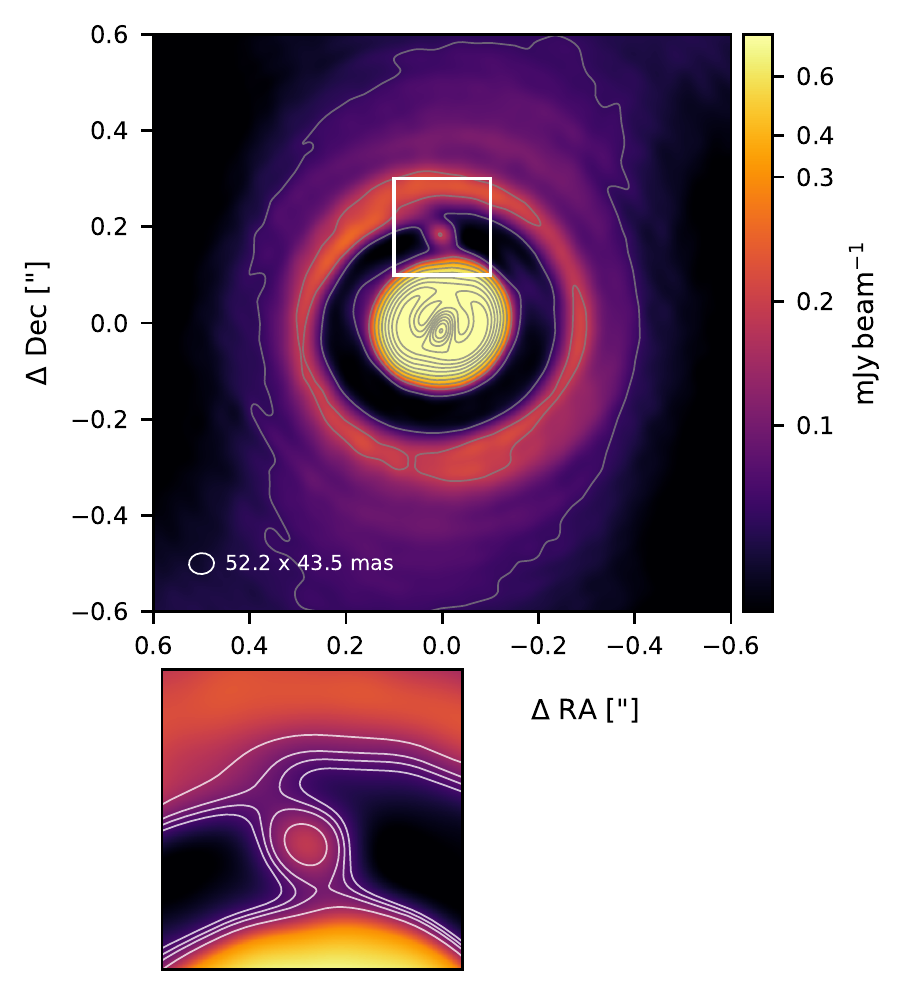}
	\caption{ALMA Band 6 synthetic observation at 1.3 mm of the dust continuum. The image was obtained post-processing an earlier dump compared to the one used in Figure~\ref{fig:sphere_alma}, over an integration time of 5 hours and removing any noise. The gray contours go from 0.03 up to 3.0 mJy beam$^{-1}$, over 20 bins. The white rectangle indicates the region shown in the inset. The contours in the inset corresponds to 0.05, 0.075, 0.1 and 0.15 mJy~beam$^{-1}$.}
	\label{fig:cpd_emission}
\end{figure}

Finally, we compute the dust temperature structure using MCFOST and produce a continuum image at 1.3~mm (240~GHz). In contrast to the IR scattered light observations, much of the disc is optically thin in mm continuum emission, so these observations primarily trace the disc mid-plane. The lower left panel of Figure~\ref{fig:sphere_alma} shows a synthetic Band 6 ALMA map, calculated using the CASA software \citep{2007ASPC..376..127M}. The 10~M$_{\rm Jup}$ embedded planet at 32~au is massive enough to open a gap in the dust. The large annular dust gap is detected in the sub-mm observations and can also be seen in our simulations (see lower right panel of Figure~\ref{fig:sphere_alma}). 
We additionally compute a Band 6 synthetic observation of a snapshot from the hydrodynamic simulation at about 1300~yr, using an integration time of 5 hours and removing any noise. Figure~\ref{fig:cpd_emission} shows the post-processed 1.3~mm dust emission where a circumplanetary disc is visible. The inset illustrates a zoom in on the circumplanetary disc itself, with the contours indicating the emission at 0.05, 0.75, 0.1 and 0.15 mJy~beam$^{-1}$.

\section{Discussion}
\label{sec:discussion}
We aimed to provide a dynamical explanation for the substructures in the \hd{} disc. We have shown that a central misaligned binary and an embedded planetary companion is the minimal configuration necessary.

If confirmed, this would be the first known example of a circumbinary planet that is significantly misaligned to the binary orbital plane. \cite{2021AJ....161..146J} have investigated the possible presence of a planetary companion within the gap of \hd{}. Using high contrast direct imaging observations carried out with NaCo/VLT, they were however unable to resolve any Jupiter mass planet below 50~au, at the distance of Upper Sco.

The HD~98800 system hosts a protoplanetary disc that orbits a binary in a polar configuration \citep{2019NatAs...3..230K}, stable enough to form misaligned planets. However, no planets have been currently detected within HD~98800. To date, 14 transiting circumbinary planets are known, and they are all co-planar, within a few degrees, with the binary orbit \citep{2020AJ....159..253K}. The frequency of aligned planetary companions is consistent with the occurrence rate of equivalent planets around single stars. If, however, inclined circumbinary planets are common, the inferred occurrence rate increases substantially \citep{2014MNRAS.444.1873A}. 
Detecting circumbinary planets through the transit method is however very challenging. It is particularly hard to obtain a regular transit where the planet passes in front of both stars \citep{2014A&A...570A..91M}. 

HD143006 adds to a growing collection of discs suggested to contain misaligned central binaries, including HD~142527 \citep{2016A&A...590A..90L, 2018MNRAS.477.1270P}, GW~Ori \citep{2020Sci...369.1233K}, AB~Aur \citep{2019MNRAS.489.2204P,2020MNRAS.496.2362P}, and HD~98800 \citep{2019NatAs...3..230K} to name a few.

\subsection{Misaligned circumplanetary disc}
During giant planet formation, a sub-disc forms around the Jovian-mass planet. Such circumplanetary discs regulate the accretion onto the planet and are the origin of satellites \citep[e.g.][]{2014ApJ...782...65S, 2017MNRAS.464.3158S}. 
Our hydrodynamic simulations of \hd{} show the formation of a circumplanetary disc around the 10 Jupiter mass planet. This disc is not co-planar with the orbit of the planet, or with the outer disc. The interaction with the misaligned inner binary, and possibly the misaligned inner disc, produces a significant tilt of the circumplanetary disc. Even with a 10~M$_{\rm Jup}$ planet, the disc around such a planet is unlikely to have been identified in existing observations as the detection of circumplanetary discs requires deep observations at high resolution \citep{2018MNRAS.473.3573S, 2018MNRAS.479.1850Z}. In the case of \hd{}, the ALMA observations are limited by the telescope resolution and integration time \citep{2018ApJ...869L..50P}, therefore, if existent, such circumplanetary disc would have remained undetected. 
We derived an estimate of the flux of the circumplanetary disc and find that the emission peaks at $\sim$~0.1 -- 0.15 mJy~beam$^{-1}$. This result shows that under ideal circumstances (i.e. no noise, 5 hours of integration time) the circumplanetary disc may be bright enough to be detectable.

Recently, \cite{2020ApJ...898L..26M} investigated the conditions under which the tilt of a circumplanetary disc can grow, arguing that a rapid growth is typically favoured. This could potentially have implications for both satellites formation and additionally the observability of circumplanetary discs.

\subsection{Observational predictions}

\subsubsection{Disc induced features}
We can calculate the precession time-scale for both the inner and outer discs. As shown in Section~\ref{sec:case3}, we find that the inner disc precesses $\sim10^2$ times faster than the material at larger radii. Assuming that the inner disc is still misaligned to the binary, this implies that the observed narrow-lane shadows should precess on a time-scale much shorter than the over-brightness detected at larger radii. The apparent motion of the shadow cast by the precessing inner disc is complex, and depends sensitively on the (unknown) orbital phases \citep{2020MNRAS.493L.143N}. However, the inner disc precession rate of $\sim 0.3^{\circ}$~yr$^{-1}$ suggests this could potentially be detected on a time-scale of $\lesssim10$~yr. 
Eventually the inner disc will stop precessing and align with the binary. Under these circumstances, the produced shadows will not move.

\subsubsection{Planet induced features}
The orbital period of the planet embedded at 32~au is approximately 135 years, 10 times faster than the precession timescale for the inner disc. This means that the planet advances on its orbit by $\sim 3^{\circ}$~yr$^{-1}$. Therefore, when looking at observational predictions for the planet we expect to see a variation on a relatively short time. 
The resolution of the current CO observations corresponds to $\sim$10--15 degrees of orbital motion at the planet`s position. Changes to the perturbation (kink) in the channel maps may therefore be detectable within 5--10 years. 

Although weakly shown in Figure~\ref{fig:sphere_alma}, we additionally detect a shadow induced by the planet in the scattered light image. The planet and its disc are indeed able to cast a small shadow onto the outer disc, at the edge of the gap. The time variability of this illumination effect is again relatively short and, considering that scattered light observations have higher resolution, we expect to be able to detect any significant change within the next few years.

\subsection{Caveats}
Our model successfully reproduces the major features in the scattered light observations and the CO kinematics, but does not match some of the mm continuum features seen with ALMA. In particular, the formation of a secondary shallow gap at larger radii \citep[$\sim$~51~au,][]{2018ApJ...869L..50P} is difficult to achieve with a single planet located at 32~au. \cite{2017ApJ...843..127D} showed that a super-Earth is able to open a double gap and such gaps are expected to be shallower. However, this mechanism requires a much lower disc viscosity ($\alpha \lesssim 10^{-4}$) than is used here. 
Another scenario that does not require an additional body, is offered by the work presented in \cite{2020MNRAS.492.3306A}. They have studied the formation of a warp induced by differential precession of a gas+dust disc around a binary. Due to the relative velocity between gas and dust, dust particles pile up at the warp location and generate dusty rings. Alternatively, a second smaller planet located at larger radii is in principle able to open a shallower gap, despite not being detected in the CO emission. Nonetheless, adding an extra body would result in an even more complicated scenario, and the long-term stability of such a system is not clear.

The bright arc in the south-east direction is also hard to reproduce. When considering Figure~1 in \cite{2018ApJ...869L..50P} the peak of the asymmetric structure at $\sim$~74~au is bright enough to require dust trapping. 
They model this arc using a prescription for a dust vortex, but this interpretation again requires very low viscosity \citep[e.g][]{2009A&A...493.1125L, 2014ApJ...795...53Z}. Future work using a lower viscosity than we have employed here may resolve this discrepancy.
Our simulations are also limited by the radial extent of the disc, which is initially set to 100~au. The choice of outer radius influences the development and appearance of sub-structures in the outer disc. For our chosen outer radius and computational time of our simulation, the outer disc is still yet to reach a steady state. This potentially explains the lack of substructure at larger radii. 

Despite these limitations, our model is able to correctly predict the formation of a clear annular gap. If we want to be more precise, the simulations slightly over-predict the radial width of the gap, as shown in the bottom panels of Fig.~\ref{fig:sphere_alma}. A more extensive study exploring a lower mass planet, a higher disc viscosity and/or a higher disc aspect ratio \citep{2006Icar..181..587C}, would be needed to help minimize the gap width and obtain a closer match with the observation.

We stress that we have not explored a large range of model parameters in order to find the closest match to the observations. Instead this study is a proof of concept. Our results show that a relatively simple model -- an inclined binary and an embedded planet -- can explain the key observed sub-structures in \hd{}.

\section{Conclusions}
\label{sec:concs}
We investigate three possible configurations that could explain the sub-structures observed in \hd{}. Using 3D hydrodynamic simulations we model the disc and computing synthetic observations we compare our results with the current data \citep{2018A&A...619A.171B, 2018ApJ...869L..41A, 2018ApJ...869L..50P}.

We first considered an inclined binary, which breaks and tilts the disc.  
This configuration can reproduce part of the features detected in \hd{}, specifically the shadowing effects \ref{itm:shadows}, seen in scattered light (see Section~\ref{sec:intro}). The misaligned binary can also drive a relative misalignment which matches the one measured from the observations \ref{itm:relmis}. This model, however, fails in reproducing both the annular dust gap observed in the mm continuum and the kink seen in the channel maps. 
In fact, the mm observation and CO channel maps suggest the presence of a planet further out in the disc. We then examined an embedded planetary companion orbiting on an inclined orbit around a single star. When looking at the kinematics, even a small planet leaves a trace detectable in the channel maps \ref{itm:kink}. However, the misaligned planet is massive enough to break the disc in two and to misalign the disc material interior to its orbit. The newly formed misaligned inner disc produces a perturbation in the gas velocity pattern that is not detected in the channel maps.

We therefore propose a third possibility, involving both a misaligned central binary and an embedded planet aligned with the outer disc. 
Our model reproduces the main features inside 40~au in the scattered light image, channel maps and ALMA observations. We thus conclude that the minimum configuration for \hd{} requires an inner, misaligned binary and a planet co-planar with the disc. We suggest $\sim$3--10~M$_{\rm Jup}$ planet located at 32~au within the dust gap. Additionally, our study predicts the presence of a kink at the planet's location. This result confirms previous findings by both \cite{2018ApJ...869L..50P} and \cite{2020ApJ...890L...9P}, and helps constraining the mass and location of the planet in \hd{}. 
Current direct imaging observations are not sensitive to the proposed planet \citep{2021AJ....161..146J}.
Future observations at higher angular resolution (i.e. with SPHERE or MUSE) might be able to detect such a massive planet and ultimately verify our findings. If confirmed, to our knowledge this would be the first case of a misaligned circumbinary planet.

\section*{Acknowledgments}
We thank the anonymous referee for their constructive review that improved the clarity of the paper. We also thank Rebecca Martin for fruitful discussions.
GB acknowledges support from the University of Leicester through a College of Science and Engineering PhD studentship. GB is also supported by a Royal Society Dorothy Hodgkin Fellowship. RN acknowledges support from UKRI/EPSRC through a Stephen Hawking Fellowship (EP/T017287/1). 
RDA has received funding from the European Research Council (ERC) under the European Union's Horizon 2020 research and innovation program (grant agreement No 681601). 
NC has received funding from the European Union's Horizon 2020 research and innovation programme under the Marie Sk\l odowska-Curie grant agreement No 210021. 
This project has been carried out as part of the European Union's Horizon 2020 research and innovation programme under the Marie Sk\l odowska-Curie grant agreement No 823823 (DUSTBUSTERS). DP and CP acknowledge funding from the Australian Research Council via FT130100034, FT170100040 and DP180104235.
This research used the DiRAC \textit{Complexity} system, operated by the University of Leicester IT Services, which forms part of the STFC DiRAC HPC Facility. Figure~\ref{fig:density_render} was made using SPLASH \citep{splash}, while Figures~\ref{fig:sphere_alma} \& \ref{fig:kinematics} were made with pymcfost and matplotlib \citep{Hunter:2007}. 


\section*{Data Availability}
The data underlying this article will be shared on reasonable request to the corresponding author. The code \phantomcode{} used in this work is publicly available at \url{https://github.com/danieljprice/phantom}, as well as the code MCFOST, which is available upon request. The software CASA is also public and can be downloaded from \url{https://casa.nrao.edu/casa_obtaining.shtml}.



\bibliographystyle{mnras}
\bibliography{mybibliography.bib}




\appendix

\section{Temperature structure} \label{tempappendix}

\begin{figure}
	\centering
	\includegraphics[width=\columnwidth,trim={0cm 0cm 0cm 0cm},clip]{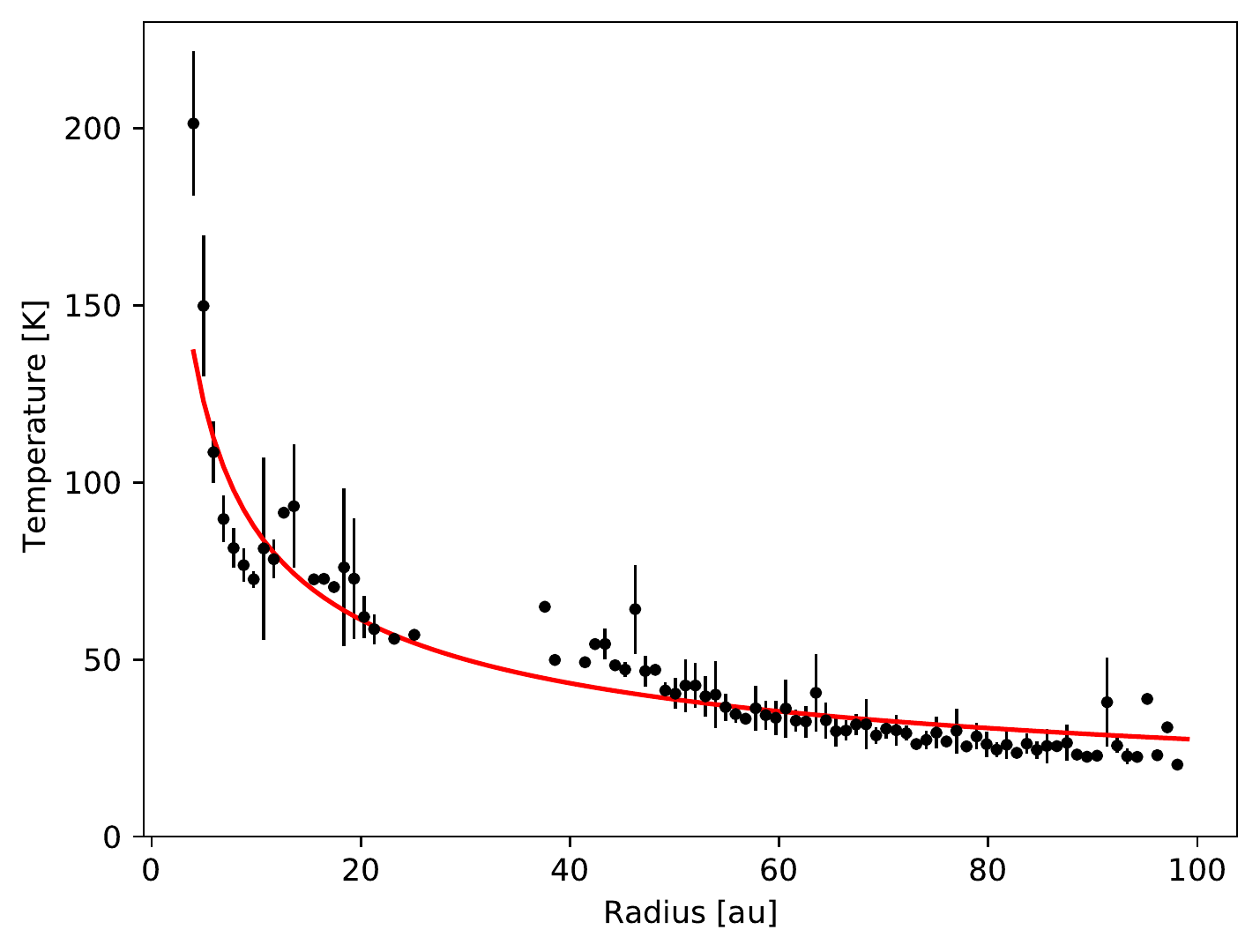}
	\caption{Comparison between the mid-plane temperature structure calculated by MCFOST, used to produce the synthetic images (black points, with error bars indicating the standard deviation), and the vertically isothermal approximation adopted in SPH (red line).}
	\label{fig:temp_midplane}
\end{figure}

\cite{2019MNRAS.484.4951N} have previously investigated a possible discrepancy between the adopted radial temperature profile and the vertical scale-height derived in MCFOST. We perform the same analysis, where we extract and average the temperature of those Voronoi cells that are within $|z| < 0.03$ from the warped mid-plane. 
The temperature obtained from the MCFOST calculations are consistent with the locally isothermal approximation used in the hydrodynamic simulations to within the error bars across most of the radial extent of the disc, as shown in Figure~\ref{fig:temp_midplane}. We note that in the gap ($\sim 32 \pm 5$ au) the gas density is negligible, so we do not plot the temperature in this region.

\bsp	
\label{lastpage}
\end{document}